\documentstyle[epsf,aps,12pt]{revtex}
\title{Bottom-Strange Associated Production at High Energy  $e^{+}e^{-}
$ Colliders in Standard Model}
\author{
\vglue 1cm
{\large Chao-Shang Huang $^a$, Xiao-Hong Wu $^a$ and Shou-Hua Zhu $^{b,a}$} \\
\vglue 0.5cm
$^a$ Institute of Theoretical Physics, Academia Sinica, P. O. Box
2735,\\ Beijing 100080, P. R. China \\
$^b$ CCAST (World Lab), P.O. Box 8730, Beijing 100080, P.R. China \\
\vglue 0.5cm
 }
\date{\today}
\begin{document}
\maketitle
\vskip 2.5cm
\begin{minipage}{15cm}
\begin{center} Abstract \end{center}
We investigate the flavor changing neutral current bsV(V=$\gamma$,Z) couplings
in the production vertex  for the process $e^+e^-\rightarrow b\bar s
\mbox{ or } \bar b s$
in the standard model. The precise calculations keeping all
quark masses non-zero are carried out.
Production cross sections are found to be the order of $10^{-3}$ fb at LEP II
and the order of $10^{-1}$ fb when center-of-mass energy is near the mass of
neutral gauge boson Z. \end{minipage}
\newpage

\section{Introduction}

\hspace{4mm}There are no flavor changing neutral currents (FCNC) at
tree-level in the standard model (SM). FCNC appear at loop-levels and
consequently offer a good place to test quantum effects of the
fundamental quantum field theory on which SM based. Furthermore, they are
very small at one loop-level due to the unitarity of Cabbibo-Kobayashi-Maskawa
 (CKM) matrix. In models beyond SM new particles beyond the particles in SM
may appear in the loop and have significant contributions to flavor
changing transitions. Therefore, FCNC interactions give an ideal place to
search for new physics. Any positive observation of FCNC couplings
deviated from that in SM would unambiguously signal the presence of new
physics. Searching for FCNC is clearly one of important goals of the next
generation of high energy colliders~\cite{pro}.

The flavor changing transitions involving external up-type quarks which are
due to FCNC couplings are much more suppressed than those involving
external down-type quarks in SM. The effects for external up-type quarks
are derived by virtual exchanges of down-type quarks in a loop for which
GIM mechanism ~\cite{gim} is much more effective because the mass splitting
between down-type quarks are much less than those between up-type quarks.
Therefore, for example, the bs transition which is studied in the paper
has larger probability to be observed than that for the tc transition.

The b-hadron system promises to give a fertile ground to test the SM and 
probe new physics. 
The FCNC vertices bsV(V=$\gamma$, Z) have been extensively examined in rare decays
of b-hadron system~\cite{my,neu,rev}. The observation of FCNC processes in both
the exclusive $B\rightarrow K^{\star}\gamma$ and inclusive $B\rightarrow X_s\gamma$
channels has placed the rare B decays on a new footing and has put a stringent
constraint on classes of models~\cite{hew}. Analyses of the inclusive decay $B
\rightarrow X_s l^+l^-$ show that in the minimal supergravity model(SUGRA) there are regions in the parameter space where the branching ratio of $b\rightarrow s l^{+}l^{-}(l=e, {\mu})$ is enhanced by about 50\% compared to the SM~\cite{gost}
and the first distinct signals of SUSY could come from the observation of
$B \rightarrow X_s{\mu}^{+}{\mu}^{-}$ if tan$\beta$ is large ( $\geq 30$ )
and the mass of the lightest neutral Higgs boson m$_h$ is not too large (say, less than 150 Gev)~\cite{my}. The B factories presently under construction will collect some $10^7$---$10^8$ B mesons per year which can be used to obtain good precision on low branching fraction modes.

The FCNC vertices bsV(V=$\gamma$, Z) can also be investigated via bottom-strange associated production. In the paper we shall investigate the process
\begin{eqnarray}
e^+e^-\rightarrow b\bar{s} \mbox{ or } \bar{b}s. 
\end{eqnarray}
Comparing b quark rare decays where the momentum transfer $q^2$ is
limited, i. e., it should be less or equal to mass square of b quark
$m_b^2$, the production process (1) allows the large (time-like) momentum
transfer, which is actually determined by the energies available at
$e^+e^-$ colliders. The reaction (1) has some advantages because of the
ability to probe higher dimension operators at large momenta and striking
kinematic signatures which are straightforward to detect in the clean
environment of $e^+e^-$ collisions. In particular, in some extensions of
SM which induce FCNC there are large underlying mass scales and large
momentum transfer so that these models are more naturally probed via
b$\bar{s}$ associated production than b quark rare decays. 

It has been shown that the cross sections of $e^+e^-\rightarrow t\bar c$ in SM are
too small to be observed at LEP or NLC~\cite{tc}. As pointed above, in SM the cross 
sections of $e^+e^-\rightarrow b \bar s$ should be much larger than those of
tc final states. Are they large enough to be seen at LEP or NLC? 
In the paper we would like to address the problem by calculating cross sections and
backward-forward asymmetry of the process (1) in SM. 
\section{Analytic calculations}
In SM for the process (1) there are three kinds of Feynman diagram at one
loop, self energy-type, triangle and box diagram, which are shown in Fig.1. We
carry out calculations in the Feynman-t'Hooft gauge. The
contributions of the neutral Higgs H and Goldstone bosons
$G^{0,\pm}$ which couple to electrons are neglected since they are
proportional to the electron mass and we have put the mass of
electron to zero.
\par
We do the reduction using FeynCalc \cite{3} and keep all masses non-zero except
for the mass of electron. To control
the ultraviolet divergence, the dimensional regularization is used. As a
consistent check, we found that all divergences are canceled in the sum of
contributions of all Feynman diagrams. The calculations are carried out in the
frame of the center of mass system (CMS) and  Mandelstam variables
have been employed: \begin{eqnarray}
s=(p_1 +p_2)^2 =(k_1 +k_2)^2 \hspace{7mm} t=(p_1 -k_1)^2 \hspace{7mm}
u=(p_1 -k_2)^2,
\end{eqnarray}
 where
$p_1,p_2$ are the
momentum of electron and positron respectively, and $k_1,k_2$ are the momentum of 
bottom quark $b$, and anti-strange quark $\bar{s}$ respectively.
\par
The amplitude of process $e^{+}e^{-} \rightarrow \bar b s$ can be 
expressed as

\begin{eqnarray}
M &=& \sum_{j=u,c,t} 16\pi^2 \alpha^2 V_{jb}^{\star}V_{js}[g_1 \bar{u_b} \gamma^{\mu}  P_R  v_s  \bar{v_e}  \gamma_{\mu} P_R  u_e +
 g_2 \bar{u_b}  \gamma^{\mu}  P_L v_s  \bar{v_e}  \gamma_{\mu}  P_R  u_e + 
g_3 \bar{u_b}  \gamma^{\mu}  P_R  v_s  \bar{v_e}  \gamma_{\mu}  P_L  u_e +\nonumber\\
&& g_4 \bar{u_b}  \gamma^{\mu}  P_L  v_s  \bar{v_e}  \gamma_{\mu}  P_L  u_e +
 g_5 \bar{u_b}  P_R  v_s  \bar{v_e}  \not\!{k_1}  P_R  u_e +
 g_6 \bar{u_b}  P_L  v_s  \bar{v_e}  \not\!{k_1}  P_R  u_e +
g_7 \bar{u_b}  P_R  v_s  \bar{v_e}  \not\!{k_1}  P_L  u_e +\nonumber\\
&&g_8 \bar{u_b}  P_L  v_s  \bar{v_e}  \not\!{k_1}  P_L  u_e + 
 g_9 \bar{u_b} \not\!{p_1}  P_L  v_s  \bar{v_e}  \not\!{k_1}  P_L  u_e +
 g_{10} \bar{v_e}  \gamma^{\mu}  P_L  u_e  \bar{u_b}  \gamma_{\mu}  \not\!{p_1}  P_R  v_s + \nonumber\\
&& g_{11} \bar{v_e} \gamma^{\mu}  P_L  u_e  \bar{u_b}  \gamma_{\mu}  \not\!{p_1}  P_L  v_s]
\end{eqnarray} 
where $\alpha$ is fine structure constant, $V_{ij}$ is CKM matrix element, $P_L$ is 
defined as $(1-\gamma^5)/2$, and $P_R$ is defined as $(1+\gamma^5)/2$.
The expressions of the coefficients $g_j(j=1,2,...11)$ can be
found in Appendix.
\par
Having the amplitude M, it is straightforward to obtain the differential
cross section by
\begin{eqnarray}
\frac{d\sigma}{dcos\theta} ={\frac{N_c}{16\pi}}
{\frac{|\vec{k_1}|}{s^{3\over{2}}}} {\frac{1}{4}} {\sum_{spins}|M|^2} 
\end{eqnarray}
Where $N_c$ is the color factor and $\theta$ is the angle between
incoming electron $e^{-}$ and outgoing
bottom quark $b$.

\section{Numerical results}
\hspace{4mm} In the numerical calculations the following values of the
parameters have been used ~\cite{pdg}: 
\begin{eqnarray}
m_e=0,\hspace{4mm}m_u=0.005Gev,\hspace{4mm}m_c=1.4Gev,\hspace{4mm}m_t=175Gev,\hspace{4mm}m_s=0.17Gev,\hspace{4mm}\nonumber\\
m_b=4.4Gev,\hspace{4mm}m_w=80.41Gev,\hspace{4mm}m_z=91.187Gev,\hspace{4mm} 
\Gamma_z=2.5Gev,\hspace{4mm} \alpha=\frac{1}{128} \nonumber
\end{eqnarray}
\par
In order to keep the unitary condition of CKM matrix exactly, we employ the 
standard parameterization and took the values ~\cite{pdg,ckm}
\begin{eqnarray}
s_{12}=0.220,\hspace{7mm} s_{23}=0.039,\hspace{7mm} s_{13}=0.0031,\hspace{7mm} \delta_{13}= 70^{\circ} \nonumber
\end{eqnarray}
Numerical results are shown in Figs. 2, 3, 4.
\par
In Fig.2, we show the total cross section $\sigma_{tot}$ of the process $e^{+}e^{-} \rightarrow b\bar{s}$ as a function of 
the center-of-mass energy $\sqrt{s}$.
There are three peaks, corresponding to the pole of neutral gauge boson
$Z^0$, a pair of charged gauge boson $W$ threshold, and a pair of top quark 
$t\bar{t}$ threshold respectively.
In  most of high energy region, total cross section is the order of
$10^{-3}$ fb, which is too small to be seen at LEP II or planning NLC colliders. Therefore, even a small number of bs events, detected at LEP II or NLC, will unambiguously indicate  new FCNC couplings beyond SM. Smallness of the total cross section can easily be understood. One has
\begin{eqnarray}
\sum_{spins}|M|^2 & = & e^8 |\sum_{j=u,c,t} V_{jt}^{\star}V_{jc} f(x_j, y_j)|^2
\nonumber \\ & = &  e^8 |V_{tb}^{\star} V_{ts} \frac{m_t^2-m_c^2}{m_w^2}
\frac {\partial f}{\partial x_j}|_{x_j,y_j=0} + ...|^2,
\end{eqnarray}
due to GIM mechanism, where $x_j=m_j^2/m_w^2, y_j=m_j^2/s$, and "..." denote the less important terms for $\sqrt s\ge 200$ Gev. Assuming $\frac {\partial
f}{\partial x_j}|_{x_j,y_j=0} $ = O(1), one obtains from eqs. (4), (5)\\
$$ \sigma \sim 10^{-3} fb$$
at $\sqrt s$ = 200 Gev.
\par
We fixed the center-of-mass energy $\sqrt{s}$ at $200$ Gev. Differential
cross  section of the process at the energy as a function of $\cos{\theta}$
 is shown in Fig.3.
\par
The Fig.4 is devoted to the backward-forward asymmetry
\begin{equation}
A_{FB}=\frac{\int_{0}^{\pi/2}{d\sigma \over d\theta}d\theta -
\int_{\pi/2}^{\pi}{d\sigma \over d\theta}d\theta }
{\int_{0}^{\pi/2}{d\sigma \over d\theta}d\theta +
\int_{\pi/2}^{\pi}{d\sigma \over d\theta}d\theta }
\end{equation}
\par
as a function of $\sqrt{s}$.
\par 
To summarize, we have calculated the process $e^{+}e^{-} \rightarrow b\bar{s}$ in SM.  We found that the total cross 
section is of the order of $10^{-3} fb$ in the
 high energy region which is still too small to be seen at LEP II or planning NLC. However, it is worth to note that
 the total cross section at Z resonance may reach as large as $10^{-1}$ fb. Therefore, it is possible to see the process if a luminosity reaches 100-1000 $fb^{-1}$. In addition to that, the process is
 of  a good place to search for new physics.

\section*{ Acknowledgments}

This research was supported in part by the National Nature Science
Foundation of China and the post doctoral foundation
of China. S.H. Zhu gratefully acknowledges the
support of  K.C. Wong Education Foundation, Hong Kong.

\section*{Appendix}
\begin{eqnarray}
g_{1} &=&  m_s(B_0^a (m_b^2 - m_s^2) (m_j^2 - m_w^2) (m_j^2 + 2 m_w^2) + B_0^b m_s^2 (m_b^4 - 2 m_b^2 m_j^2 + m_j^4 + m_b^2 m_w^2 + m_j^2 m_w^2 -\nonumber\\
&& 2 m_w^4) - B_0^c m_b^2 (m_s^4 - 2 m_s^2 m_j^2 + m_j^4 + m_s^2 m_w^2 + m_j^2 m_w^2 - 2 m_w^4)) (a_1 - 4 a_2 s_w^4) + \nonumber\\
&& 2 m_b m_s (2 C_{00}^e + C_{11}^e m_b^2 + C_0^e m_j^2 + C_1^e (m_b^2 + m_j^2 - 2 m_w^2) + C_{22}^e s + C_2^e s +\nonumber\\
&& C_{12}^e (m_b^2 - m_s^2 + s)) (a_3 + 6 a_4 s_w^2 - 8 a_4 s_w^4) - 6 C_{00}^d m_b m_s (a_3 + 4 a_4 s_w^2 (c_w^2 - s_w^2)) +\nonumber\\
&& 12 m_b m_s m_w^2 (C_0^d + C_1^d) (a_3 + 6 a_4 c_w^2 s_w^2 - 2 a_4 s_w^4) 
\end{eqnarray}
\begin{eqnarray}
g_{2} &=&  -a_3 m_j^2  + 8 a_4 m_j^2 s_w^4  - 6 m_w^2(C_{11}^d m_b^2 + C_{22}^d s + C_{12}^d (m_b^2 - m_s^2 + s) ) (a_3 + 8 a_4 c_w^2 s_w^2)  + \nonumber\\
&& m_b m_s^2 ( B_0^a (m_b^2 - m_s^2) (m_j^2 - m_w^2) + (m_b^2 m_s^2 - m_b^2 m_j^2 - m_s^2 m_j^2 + m_j^4  + m_j^2 m_w^2 -\nonumber\\
&& 2 m_w^4) (B_0^b -B_0^c) + m_w^2 (B_0^b (2 m_b^2 -m_s^2) -B_0^c (2 m_s^2 - m_b^2) )) (a_1 + 6 a_2 s_w^2 - 4 a_2 s_w^4) - \nonumber\\
&& 6 C_2^d m_w^2 (a_3 (m_b^2 - m_s^2 + s) + 4 a_4 s_w^2 ((m_b^2 - m_s^2) (c_w^2 - s_w^2) + 2 s c_w^2) ) - 6 C_{00}^d (a_3 (m_j^2 + 6 m_w^2) + \nonumber\\
&& 4 a_4 s_w^2 (c_w^2 m_j^2 + 12 c_w^2 m_w^2 - m_j^2 s_w^2)) + 2 (2 C_{00}^e + C_{11}^e m_b^2 + C_{22}^e s + C_{12}^e (m_b^2 - m_s^2 + s) )\nonumber\\
&& (a_3 (m_j^2 + 2 m_w^2) + 4 a_4 s_w^2 (3 m_w^2 - 2 m_j^2 s_w^2 - 4 m_w^2 s_w^2)) + 2 C_0^e m_j^2 (a_3 (m_b^2 + m_s^2 - m_j^2 - 2 m_w^2) +\nonumber\\
&&  2 a_4 s_w^2 (3 m_s^2 - 3 m_j^2 - 4 m_b^2 s_w^2 - 4 m_s^2 s_w^2 + 4 m_j^2 s_w^2 + 8 m_w^2 s_w^2)) + 2 C_2^e (a_3 s (m_j^2 + 2 m_w^2) -\nonumber\\
&& 2 a_4 s_w^2 (3 m_b^2 m_j^2 - 3 m_s^2 m_j^2 - 6 m_w^2 s + 4 m_j^2 s s_w^2 + 8 m_w^2 s s_w^2) )  - 6 C_1^d m_w^2 (a_3 (m_b^2 - m_s^2 + s) +\nonumber\\
&& 4 a_4 s_w^2 (m_b^2 c_w^2 - m_b^2 s_w^2  - 2 c_w^2 m_s^2 + 2 c_w^2 s)) - 6 C_0^d m_w^2 (a_3 (s - m_s^2 - m_j^2) +\nonumber\\
&& 4 a_4 s_w^2 (2 s c_w^2 - m_b^2 + 2 m_j^2 s_w^2  - 2 c_w^2 m_s^2)) + 2 C_1^e (a_3 (2 s m_w^2 + m_b^2 m_s^2 + m_b^2 m_j^2 - 2 m_w^2 m_s^2) + \nonumber\\
&& 2 a_4 s_w^2 (3 m_b^2 m_s^2 - 3 m_b^2 m_j^2 + 6 s m_w^2 - 4 m_b^2 m_s^2 s_w^2 - 4 m_b^2 m_j^2 s_w^2 - 8 s m_w^2 s_w^2 -\nonumber\\
&& 6 m_w^2 m_s^2 + 8 m_w^2 m_s^2 s_w^2))
\end{eqnarray}
\begin{eqnarray}
g_{3} &=&  - a_5 m_b m_s (2 D_{23}^f + D_3^f) + m_s (B_0^a (m_b^2 - m_s^2) (m_j^2 - m_w^2) (m_j^2 + 2 m_w^2) + B_0^b m_s^2 (m_b^4 - 2 m_b^2 m_j^2 +\nonumber\\
&& m_j^4 + m_b^2 m_w^2 + m_j^2 m_w^2 - 2 m_w^4) - B_0^c m_b^2 (m_s^4 - 2 m_s^2 m_j^2 + m_j^4 + m_s^2 m_w^2 + m_j^2 m_w^2 - 2 m_w^4) ) (a_1 +\nonumber\\
&& 2 a_2 s_w^2 - 4 a_2 s_w^4) + 2 m_b m_s (2 C_{00}^e + C_{11}^e m_b^2 + C_0^e m_j^2 + C_1^e (m_b^2 + m_j^2 - 2 m_w^2) + C_{22}^e s + C_2^e s +\nonumber\\
&& C_{12}^e (m_b^2 - m_s^2 + s) ) (a_3 - 3 a_4 + 10 a_4 s_w^2 - 8 a_4 s_w^4) - 6 C_{00}^d m_b m_s (a_3 - 2 a_4 (1 - 2 s_w^2)^2) +\nonumber\\
&& 12 m_b m_s m_w^2 (C_0^d + C_1^d) (a_3 - a_4 (3 - 4 s_w^2) (1 - 2 s_w^2)) 
\end{eqnarray}
\begin{eqnarray}
g_{4} &=&  - a_5 (2 D_{00}^f - (2 D_{13}^f + D_1^f) (m_b^2 - t) + 2 D_{23}^f t + D_3^f t) - a_3 m_j^2  - 4 a_4 m_j^2 s_w^2 (1 - 2 s_w^2)  -\nonumber\\
&& 6 m_w^2 (C_{11}^d m_b^2 + C_{22}^d s + C_{12}^d (m_b^2 - m_s^2 + s) ) (a_3 - 4 a_4 c_w^2 (1 - 2 s_w^2) ) + \nonumber\\
&& m_b m_s^2 (B_0^a (m_b^2 - m_s^2) (m_j^2 - m_w^2) + B_0^b (m_b^2 m_s^2 - m_b^2 m_j^2 - m_s^2 m_j^2 + m_j^4 + 2 m_b^2 m_w^2 - m_s^2 m_w^2 +\nonumber\\
&& m_j^2 m_w^2 - 2 m_w^4) - B_0^c (m_b^2 m_s^2 - m_b^2 m_j^2 - m_s^2 m_j^2 + m_j^4 - m_b^2 m_w^2 + 2 m_s^2 m_w^2 + m_j^2 m_w^2 - 2 m_w^4))\times \nonumber\\
&& (a_1 - 3 a_2 + 8 a_2 s_w^2 - 4 a_2 s_w^4) - 6 C_2^d m_w^2 (a_3 (m_b^2 - m_s^2 + s) + 2 a_4 (c_w^2 m_s^2 - c_w^2 m_b^2 - 2 c_w^2 s +\nonumber\\
&& m_b^2 s_w^2 + 2 c_w^2 m_b^2 s_w^2 - m_s^2 s_w^2 - 2 c_w^2 m_s^2 s_w^2 + 4 c_w^2 s s_w^2 - 2 m_b^2 s_w^4 + 2 m_s^2 s_w^4) ) - 6 C_{00}^d (a_3 (m_j^2 + 6 m_w^2) +\nonumber\\
&& 2 a_4 (m_j^2 s_w^2 - c_w^2 m_j^2 - 12 c_w^2 m_w^2 + 2 c_w^2 m_j^2 s_w^2 + 24 c_w^2 m_w^2 s_w^2 - 2 m_j^2 s_w^4) ) +\nonumber\\
&& 2 (2 C_{00}^e + C_{11}^e m_b^2 + C_{22}^e s)(a_3 (m_j^2 + 2 m_w^2) + 2 a_4 (2 m_j^2 s_w^2 - 3 m_w^2 + 10 m_w^2 s_w^2 - 4 m_j^2 s_w^4 - 8 m_w^2 s_w^4) ) +\nonumber\\
&& 2 C_0^e m_j^2 (a_3 (m_b^2 + m_s^2 - m_j^2 - 2 m_w^2) + a_4 (3 m_j^2 - 3 m_s^2 + 4 m_b^2 s_w^2 + 10 m_s^2 s_w^2 - 10 m_j^2 s_w^2 - 8 m_w^2 s_w^2 -\nonumber\\
&& 8 m_b^2 s_w^4 - 8 m_s^2 s_w^4 + 8 m_j^2 s_w^4 + 16 m_w^2 s_w^4) ) + 2 C_2^e (a_3 s (m_j^2 + 2 m_w^2) + a_4 (3 m_b^2 m_j^2 - 3 m_s^2 m_j^2 -\nonumber\\
&& 6 m_w^2 s - 6 m_b^2 m_j^2 s_w^2 + 6 m_s^2 m_j^2 s_w^2 + 4 m_j^2 s s_w^2 + 20 m_w^2 s s_w^2 - 8 m_j^2 s s_w^4 - 16 m_w^2 s s_w^4) ) + \nonumber\\
&& 2 C_{12}^e (m_b^2 - m_s^2 + s) (a_3 (m_j^2 + 2 m_w^2) - 2 a_4 (3 m_w^2 - 2 m_j^2 s_w^2 - 10 m_w^2 s_w^2 + 4 m_j^2 s_w^4 + 8 m_w^2 s_w^4) ) -\nonumber\\
&& 6 C_1^d m_w^2 (a_3 (m_b^2 -m_s^2 + s) + 2 a_4 (m_b^2 s_w^2 - 3 c_w^2 m_b^2 + 6 c_w^2 m_b^2 s_w^2 - 2 m_b^2 s_w^4 + 2 c_w^2 t - 4 c_w^2 s_w^2 t +\nonumber\\
&& 2 c_w^2 u - 4 c_w^2 s_w^2 u) ) - 6 C_0^d m_w^2 (a_3 (s - m_s^2 - m_j^2) + 2 a_4 (m_b^2 s_w^2 + c_w^2 m_b^2 - 2 c_w^2 m_b^2 s_w^2 - 2 m_j^2 s_w^2 -\nonumber\\
&& 2 m_b^2 s_w^4 + 4 m_j^2 s_w^4 + 2 m_s^2 c_w^2 - 4 m_s^2 c_w^2 s_w^2 - 2 s c_w^2 + 4 s c_w^2 s_w^2) ) + 2 C_1^e (a_3 (m_b^2 m_s^2 + m_b^2 m_j^2 +\nonumber\\
&& 2 m_b^2 m_w^2 - 2 m_w^2 t - 2 m_w^2 u) + a_4 (3 m_b^2 m_j^2 - 3 m_b^2 m_s^2 + 6 m_s^2 m_w^2 + 10 m_b^2 m_s^2 s_w^2 - 2 m_b^2 m_j^2 s_w^2 -\nonumber\\
&& 20 m_s^2 m_w^2 s_w^2 - 8 m_b^2 m_s^2 s_w^4 - 8 m_b^2 m_j^2 s_w^4 + 16 m_s^2 m_w^2 s_w^4 - 6 s m_w^2 + 20 s m_w^2 s_w^2 - 16 s m_w^2 s_w^4) )
\end{eqnarray}
\begin{eqnarray}
g_{5} &=&  -24 a_4 m_s s_w^2 (C_2^e m_j^2 - 2 C_2^d m_w^2) - 4 m_s (C_{11}^e m_b^2 + C_0^e m_j^2 + C_1^e (m_b^2 + m_j^2 - 2 m_w^2) ) (a_3 + 6 a_4 s_w^2 -\nonumber\\
&& 8 a_4 s_w^4) - 6 m_s (C_{11}^d m_b^2 + 2 C_0^d m_w^2 + C_1^d (m_b^2 - m_j^2 + 2 m_w^2) ) (a_3 + 4 a_4 c_w^2 s_w^2 - 4 a_4 s_w^4) - \nonumber\\
&& 6 C_{12}^d m_s (a_3 (m_b^2 - m_j^2 - 2 m_w^2) + 4 a_4 s_w^2 (m_b^2 c_w^2 - c_w^2 m_j^2 - 4 c_w^2 m_w^2 - m_b^2 s_w^2 + m_j^2 s_w^2)) -\nonumber\\
&& 4 C_{12}^e m_s (a_3 (m_b^2 - m_j^2 - 2 m_w^2) + 2 a_4 s_w^2 (3 m_b^2 - 6 m_w^2 - 4 m_b^2 s_w^2 + 4 m_j^2 s_w^2 + 8 m_w^2 s_w^2))
\end{eqnarray}
\begin{eqnarray}
g_{6} &=&  24 a_4 m_b s_w^2 (C_2^e m_j^2 - 2 C_2^d m_w^2) - 12 m_b m_w^2 (C_0^d + 2 C_1^d) (a_3 + 8 a_4 c_w^2 s_w^2) -\nonumber\\
&& 4 m_b m_j^2 (C_0^e + 2 C_1^e) (a_3 - 8 a_4 s_w^4) - 6 C_{11}^d m_b (a_3 (m_j^2 + 2 m_w^2) + 4 a_4 s_w^2 (c_w^2 m_j^2 + 4 c_w^2 m_w^2 - m_j^2 s_w^2) ) +\nonumber\\
&& 6 C_{12}^d m_b (a_3 (m_s^2 - m_j^2 - 2 m_w^2) + 4 a_4 s_w^2 (c_w^2 m_s^2 - c_w^2 m_j^2 - 4 c_w^2 m_w^2 - m_s^2 s_w^2 + m_j^2 s_w^2) ) - \nonumber\\
&& 4 C_{11}^e m_b (a_3 (m_j^2 + 2 m_w^2) + 4 a_4 s_w^2 (3 m_w^2 - 2 m_j^2 s_w^2 - 4 m_w^2 s_w^2) ) + 4 C_{12}^e m_b (a_3 (m_s^2 - m_j^2 - 2 m_w^2) +\nonumber\\
&& 2 a_4 s_w^2 (3 m_s^2 - 6 m_w^2 - 4 m_s^2 s_w^2 + 4 m_j^2 s_w^2 + 8 m_w^2 s_w^2) )
\end{eqnarray}
\begin{eqnarray}
g_{7} &=&  2 a_5 m_s (D_{23}^f + D_3^f) + 12 a_4 m_s (C_2^e m_j^2 - 2 C_2^d m_w^2) (1 - 2 s_w^2) - 4 m_s (C_{11}^e m_b^2 + C_0^e m_j^2 + C_1^e (m_b^2 +\nonumber\\
&& m_j^2 - 2 m_w^2) ) (a_3 - 3 a_4 + 10 a_4 s_w^2 - 8 a_4 s_w^4) - 6 m_s (C_{11}^d m_b^2 + 2 C_0^d m_w^2 +\nonumber\\
&& C_1^d (m_b^2 - m_j^2 + 2 m_w^2) ) (a_3 - 2 a_4 (s_w^2 - c_w^2)^2 ) - 6 C_{12}^d m_s (a_3 (m_b^2 - m_j^2  - 2 m_w^2) + 2 a_4 (c_w^2 m_j^2 -\nonumber\\
&& c_w^2 m_b^2  + 4 c_w^2 m_w^2 + m_b^2 s_w^2 + 2 c_w^2 m_b^2 s_w^2 - m_j^2 s_w^2 - 2 c_w^2 m_j^2 s_w^2 - 8 c_w^2 m_w^2 s_w^2 - 2 m_b^2 s_w^4 + 2 m_j^2 s_w^4) ) - \nonumber\\
&& 4 C_{12}^e m_s (a_3 (m_b^2 - m_j^2 - 2 m_w^2) + a_4 (6 m_w^2 - 3 m_b^2 + 10 m_b^2 s_w^2 - 4 m_j^2 s_w^2 - 20 m_w^2 s_w^2 - 8 m_b^2 s_w^4 +\nonumber\\
&& 8 m_j^2 s_w^4 + 16 m_w^2 s_w^4) )
\end{eqnarray}
\begin{eqnarray}
g_{8} &=&  -2 a_5 m_b (D_{22}^f - D_{23}^f + D_2^f) - 12 a_4 m_b (C_2^e m_j^2 - 2 C_2^d m_w^2) (1 - 2 s_w^2) - 12 m_b m_w^2 (C_0^d + 2 C_1^d) (a_3 -\nonumber\\
&& 4 a_4 c_w^2 (1 - 2 s_w^2) ) - 4 m_b m_j^2 (C_0^e + 2 C_1^e) (a_3 + 4 a_4 s_w^2 (1 - 2 s_w^2) ) - 6 C_{11}^d m_b (a_3 (m_j^2 + 2 m_w^2) -\nonumber\\
&& 2 a_4 (c_w^2 m_j^2 + 4 c_w^2 m_w^2 - m_j^2 s_w^2 - 2 c_w^2 m_j^2 s_w^2 - 8 c_w^2 m_w^2 s_w^2 + 2 m_j^2 s_w^4) ) + 6 C_{12}^d m_b (a_3 (m_s^2 - m_j^2 -\nonumber\\
&& 2 m_w^2) - 2 a_4 (c_w^2 m_s^2 - c_w^2 m_j^2 - 4 c_w^2 m_w^2 - m_s^2 s_w^2 - 2 c_w^2 m_s^2 s_w^2 + m_j^2 s_w^2 + 2 c_w^2 m_j^2 s_w^2 + 8 c_w^2 m_w^2 s_w^2 +\nonumber\\
&& 2 m_s^2 s_w^4 - 2 m_j^2 s_w^4) ) - 4 C_{11}^e m_b (a_3 (m_j^2 + 2 m_w^2) + 2 a_4 (2 m_j^2 s_w^2 - 3 m_w^2  + 10 m_w^2 s_w^2 - 4 m_j^2 s_w^4 -\nonumber\\
&& 8 m_w^2 s_w^4) ) + 4 C_{12}^e m_b (a_3 (m_s^2 - m_j^2 - 2 m_w^2) + a_4 (6 m_w^2 - 3 m_s^2 + 10 m_s^2 s_w^2 - 4 m_j^2 s_w^2 -\nonumber\\
&& 20 m_w^2 s_w^2 - 8 m_s^2 s_w^4 + 8 m_j^2 s_w^4 + 16 m_w^2 s_w^4) ) \\
g_{9} &=& 2 a_5 (D_{12}^f - 2 D_{13}^f + D_{22}^f - D_{23}^f + D_2^f) \\
g_{10} &=& - a_5 m_s (2 D_{13}^f + 2 D_{23}^f + D_3^f)  \\
g_{11} &=& a_5 m_b (2 D_{13}^f + D_1^f)
\end{eqnarray}
where $a_i$ is defined as
\begin{eqnarray}
a_1 =\frac{1}{192\pi^2sm_bm_s^2m_w^2(m_b^2 - m_s^2)s_w^2}, \hspace{3mm}a_2 =\frac{1}{768\pi^2m_bm_s^2m_w^2(m_b^2 - m_s^2)(m_z^2 -im_z\Gamma{z} - s)c_w^2s_w^4}\nonumber\\
a_3 =\frac{1}{96\pi^2sm_w^2s_w^2}, \hspace{3mm}a_4 =\frac{1}{768\pi^2m_w^2(m_z^2-im_z\Gamma{z}-s)c_w^2s_w^4},\hspace{3mm} a_5 =\frac{1}{32\pi^2s_w^4}\nonumber
\end{eqnarray}
where $c_w= cos\theta_w$ and $s_w= sin\theta_w$.
In the presentation of $g_j$ above, we have used the definition of scalar integrals 
$Bs$, $Cs$,and $Ds$\cite{3},and these
functions, $Bs$, $Cs$,and $Ds$, with superscripts a,b,...,f  have the arguments \\
\begin{math}
(0,m_j^2,m_w^2),\hspace{3mm} (m_b^2,m_j^2,m_w^2),\hspace{3mm} (m_s^2,m_j^2,m_w^2), \hspace{3mm} (m_b^2,m_s^2,s,m_w^2,m_j^2,m_w^2)\\
 \hspace{3mm} (m_b^2,m_s^2,s,m_j^2,m_w^2,m_j^2), \hspace{3cm} (0,m_b^2,m_s^2,0,t,s,0,m_w^2,m_j^2,m_w^2)\\
\end{math}
respectively. Here $m_j$ denotes the mass of up-type quark  $u,c,t$.

\newpage
\begin{figure}
\epsfxsize=12cm
\centerline{\epsffile{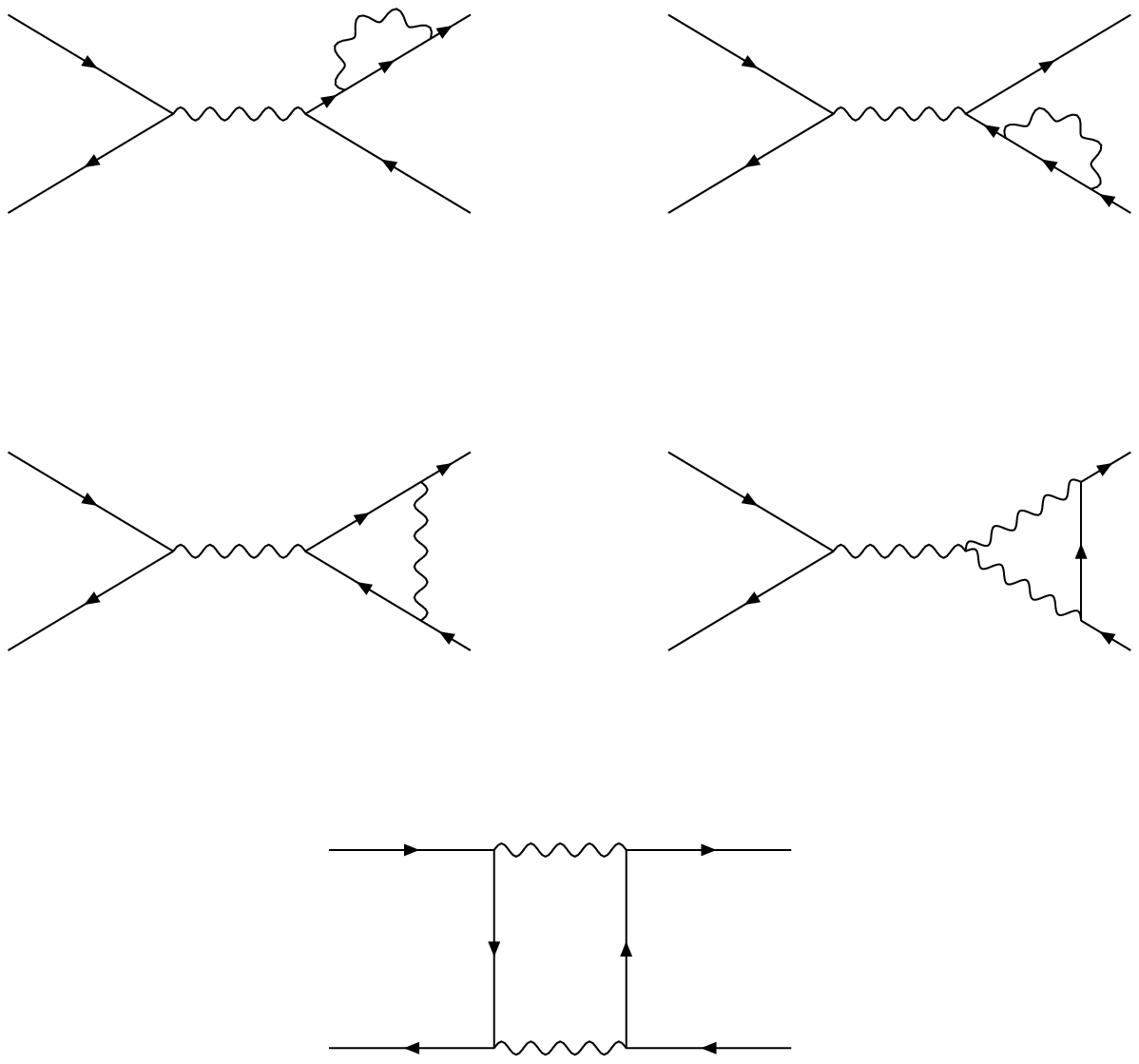}}
\caption[]{Typical Feynman diagram of
prosess $e^+e^- \rightarrow b \bar s$} 
\end{figure}

\begin{figure}
\epsfxsize=18cm
\centerline{\epsffile{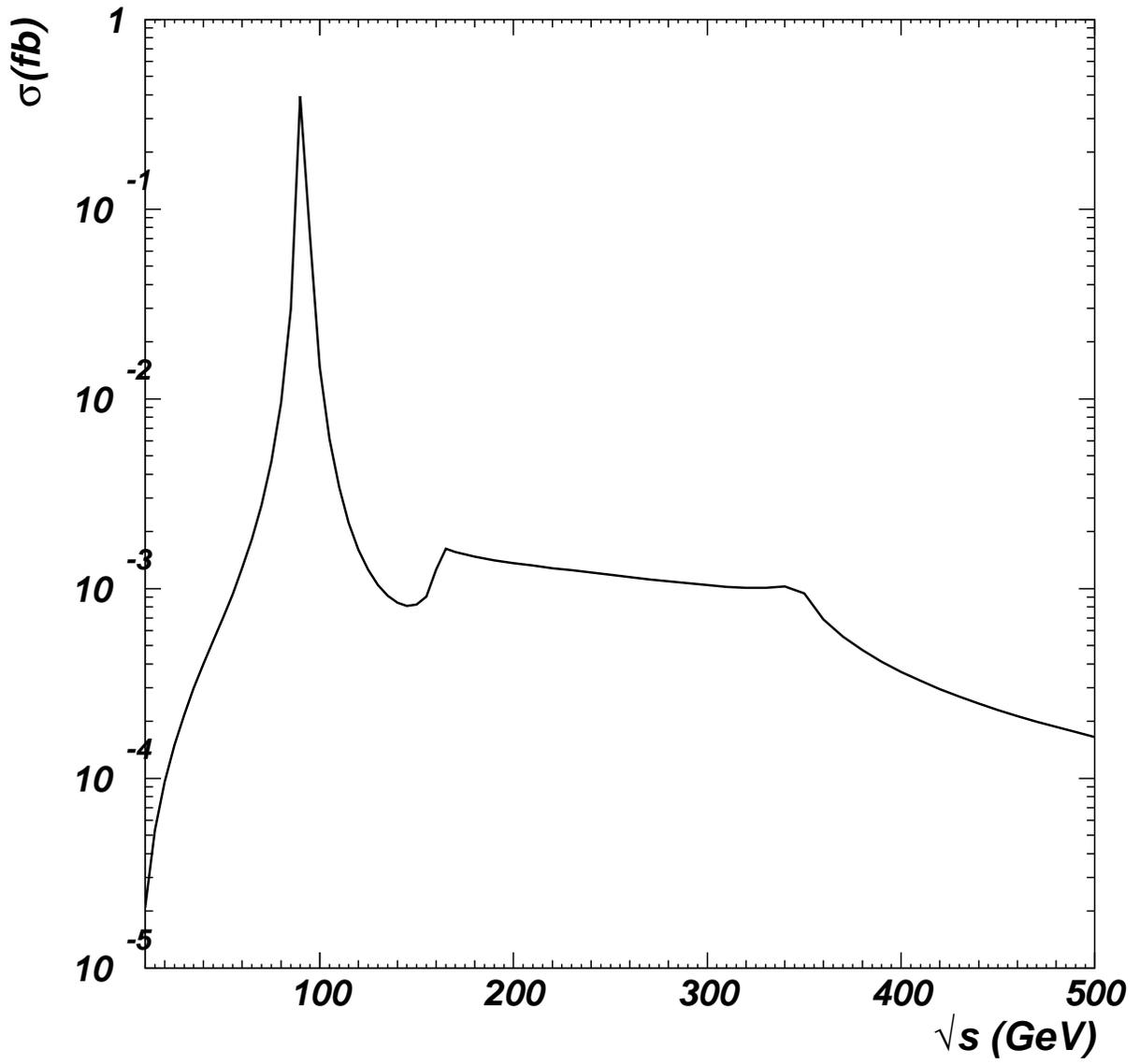}}
\caption[]{Cross section of the 
process $e^+e^- \rightarrow b \bar s$ as a function of
$\sqrt{s}$.}
\end{figure}

\begin{figure}
\epsfxsize=18cm
\centerline{\epsffile{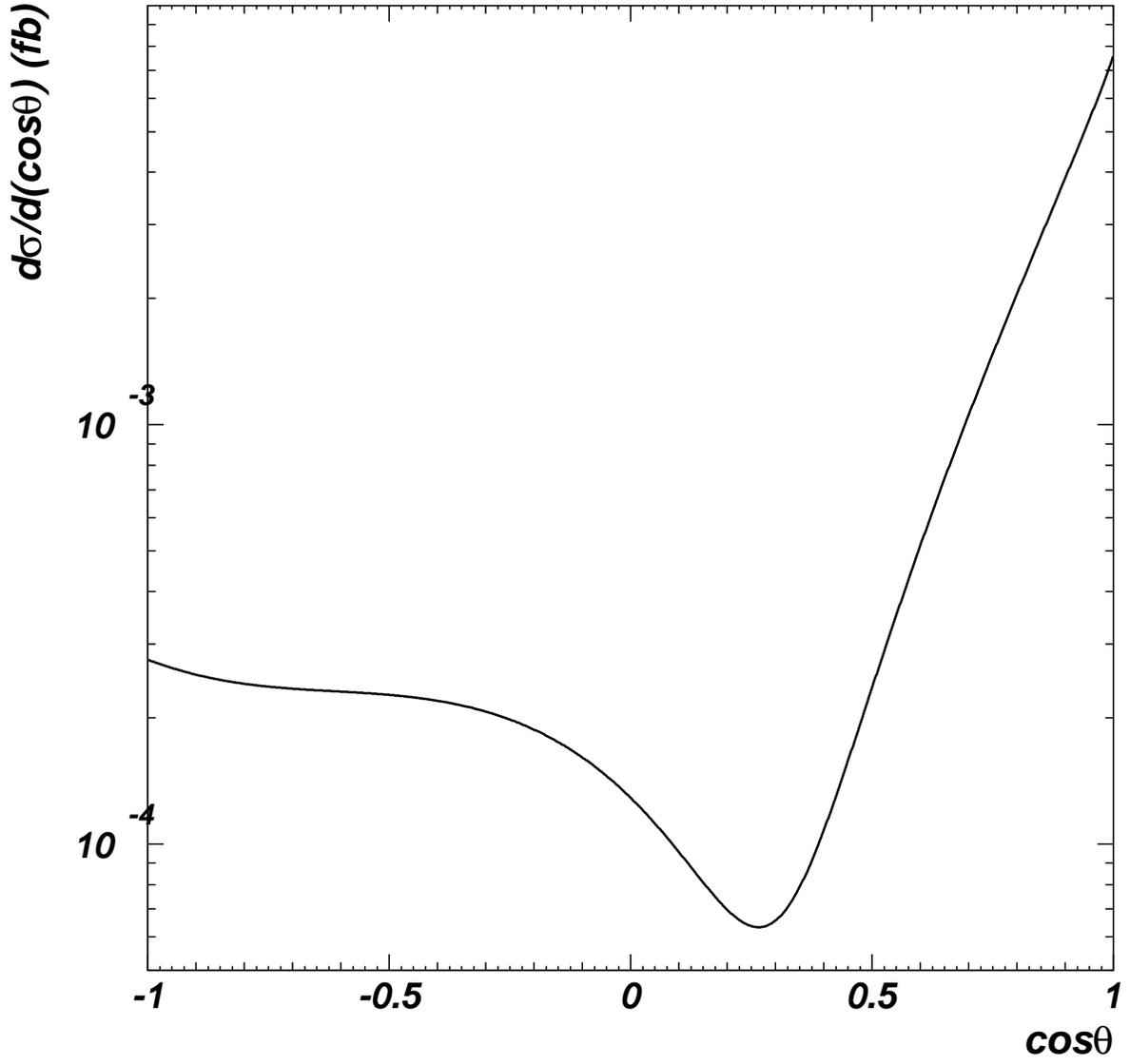}}
\caption[]{ Differential cross section of the 
process $e^+e^- \rightarrow b \bar s$, where
$\sqrt{s}=200$ GeV. 
}
\end{figure}

\begin{figure}
\epsfxsize=18cm
\centerline{\epsffile{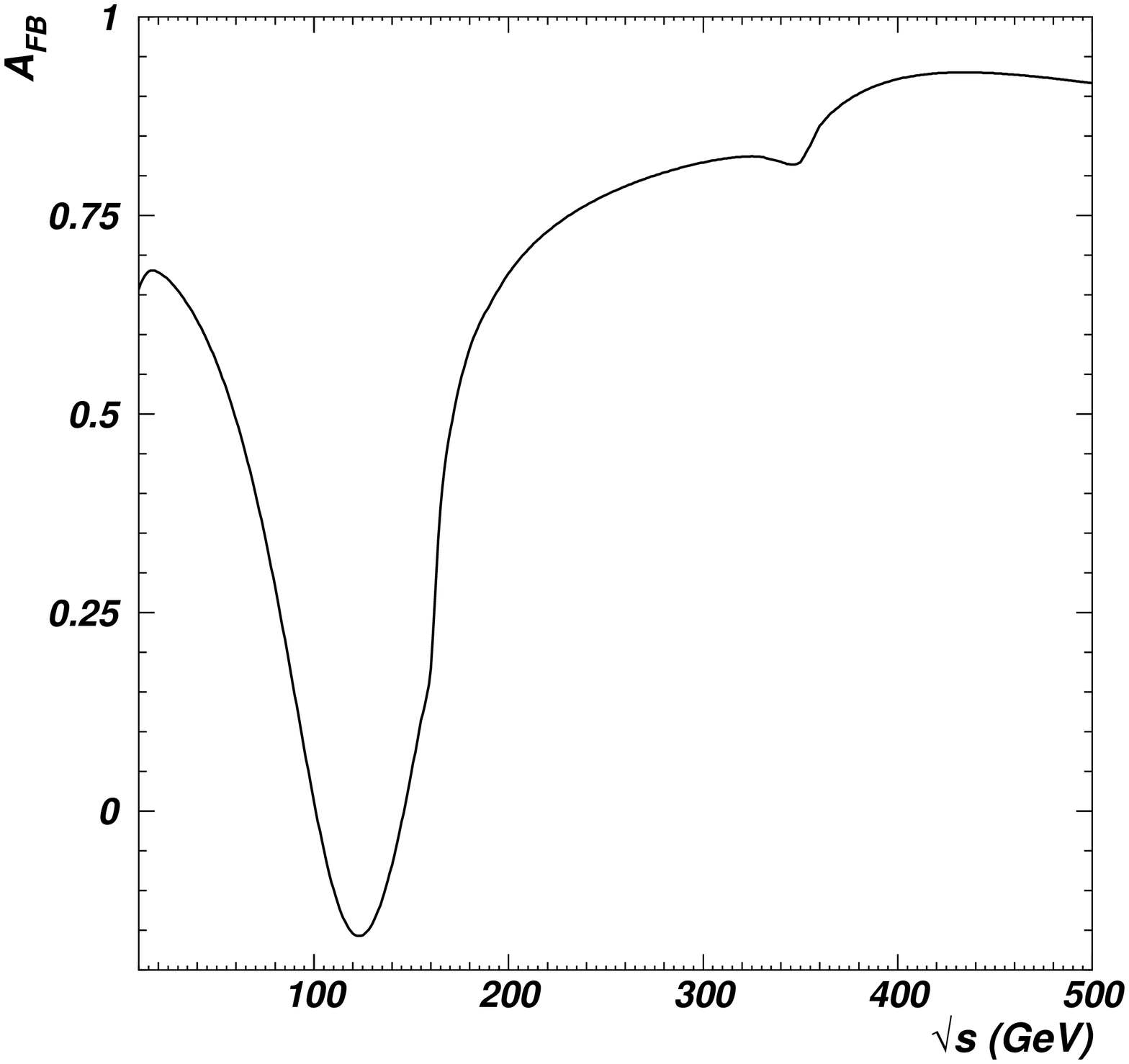}}
\caption[]{ $A_{FB}$ of the  
process $e^+e^- \rightarrow b \bar s$
as a function of $\sqrt{s}$. 
}
\end{figure}



\begin{thebibliography}{99}
\bibitem{pro}Proceedings of the Workshop on Physics and Experiments with Linear $e^+e^-$ Colliders, eds. A. Miyamoto
 and Y. Fujii, World Scientific, Singapore, 1996.
\bibitem{gim}S. L. Glashow, J. Iliopoulos, and L. Maiani, Phys. rev. {\bf D2} (1970) 1285.
\bibitem{my}Y.B. Dai, C.S. Huang and H.W. Huang, Phys. Lett. B 390 (1997) 257;
Chao-shang Huang, Wei Liao, and Qi-shu
Yan,  Phys. Rev. {\bf D59} (1998) 011701;
Chao-shang Huang and Qi-shu Yan, Phys. Lett. {\bf B442} (1998) 209.
\bibitem{neu} S. Bertolini, F. borzumati, A. Masieso and G. Ridolfi, Nucl. Phys. {\bf B353} (1991) 591;
A. Ali, G. Gindice and T. Mannel, Z. Phys. {\bf C67} (1995) 417;
P. Cho, M. Misiak and D. Wlyer, Phys. Rev. {\bf D54} (1996) 3329;
A. Masiero and L. Silvestrini, hep-ph/9711401; J. L. Hewett and
J. D. Wells, Phys. Rev. {\bf D55} (1997) 5549;
M. Ciuchini et al., hep-ph/9806308; A. L. Kagan and M. Neubert, hep-ph/9805303;
M. Neubert, hep-ph/9809377; C. S. Huang, T. Li, W. Liao, Q. S. Yan and S. H. Zhu, hep-ph/9810412;  M. Brhlik, hep-ph/9807309; H. Baer, M. Brhlik, C-H. Chen and X. Tata, Phys. Rev. {\bf D55} (1997) 4463;  J.L. Lopez, D.V. Nanopoulos, X. Wang and A. Zichichi, Phys. Rev. {\bf D51}(1995) 147;
R. Barbieri and G.F. Giudice, Phys Lett {\bf B309}(1993)86;
M.A. Diaz, Phys. Lett. {\bf B322} (1994) 591;
T. Goto and Y. Okada, Prog. of Theor. Phys. {\bf 94} (1995) 407;
R. Garisto and J. N. Ng Phys. Lett. {\bf B315} (1993) 372; B. Grinstein, M.J. Savage and M.B. Wise, Nucl. Phys. B 319 (1989) 271; 
For a recent discussion of the $b\rightarrow s\gamma$ see C. Greub and T. Hurth,
hep-ph/9809468.
\bibitem{rev}For reviews see, for example, A. Ali, hep-ph/9709507; A. J. Buras,
hep-ph/9806471; M. Misiak, S. Pokorski and J. Rosiek, hep-ph/9703442; G. Burdman,
hep-ph/9811457; J. L. Hewett, hep-ph/9803370.
\bibitem{hew}J. L. Hewett, hep-ph/9406302.
\bibitem{gost} T. Goto, Y. Okada, Y. Shimizu and M. Tanaka, Phys. Rev. D 55 (1997) 4273.
\bibitem{tc}Chao-shang Huang, Xiao-hong Wu and Shou-hua Zhu, hep-ph/9901369.
\bibitem{1}A. Axelrod, Nucl. Phys. B209,(1982)349
\bibitem{2} J.Kublbeck, M.Bohm and A. Denner, Comp. Phys. Comm, 60 (1990)165
\bibitem{pdg}Particle Data Group, Eur.Phys.J. C3,1-794(1998)
\bibitem{ckm}A. J. Buras, hep-ph/9711217, hep-ph/9806473.
\bibitem{3} G.Passarino and M.Veltman, Nucl. Phys, B160(1979) 151 ; R.Mertig, M.
Bohm, and A.Denner, Comp. Phys. Comm. 64(1991)345
\end{thebibliography}
\end{document}